\documentclass[twocolumn,amsmath,amssymb,prl]{revtex4-1}                                       

\usepackage{titlesec}
\usepackage[colorlinks,bookmarks=false,citecolor=blue,linkcolor=red,urlcolor=blue]{hyperref}
\usepackage{epsfig}
\usepackage{color}
\usepackage{subfigure}
\usepackage{graphicx}    
\usepackage{dcolumn}     
\usepackage{lipsum}      
\usepackage{braket}      
\input{insbox}

\newcommand{\be}{\begin{equation}}
\newcommand{\ee}{\end{equation}}

\definecolor{drkgr}{rgb}{0.05,0.6,0.2}

\begin{document}

\title{Sweet spot in the RuCl$_3$ magnetic system: nearly ideal $j_{\mathrm{eff}}\!=\!1/2$ moments and maximized $K/J$ ratio under pressure}

\author{Pritam Bhattacharyya}
\affiliation{Institute for Theoretical Solid State Physics, Leibniz IFW Dresden, Helmholtzstra{\ss}e~20, 01069 Dresden, Germany}

\author{Liviu Hozoi}
\affiliation{Institute for Theoretical Solid State Physics, Leibniz IFW Dresden, Helmholtzstra{\ss}e~20, 01069 Dresden, Germany}

\author{ Quirin Stahl}
\affiliation{Institut f\"ur Festk\"orper- und Materialphysik, Technische Universit\"at Dresden, 01062 Dresden, Germany}

\author{Jochen Geck}
\affiliation{Institut f\"ur Festk\"orper- und Materialphysik, Technische Universit\"at Dresden, 01062 Dresden, Germany}
\affiliation{W\"urzburg-Dresden Cluster of Excellence ct.qmat, Technische Universit\"at Dresden, 01062 Dresden, Germany}

\author{Nikolay A.~Bogdanov}
\affiliation{Max Planck Institute for Solid State Research, Heisenbergstra{\ss}e~1, 70569 Stuttgart, Germany}

\begin{abstract}
\noindent
Maximizing the ratio between Kitaev and residual Heisenberg interactions is a major goal in nowadays
research on Kitaev-Heisenberg quantum magnets.
Here we investigate Kitaev-Heisenberg exchange in a recently discovered crystalline phase of RuCl$_3$
under presure ---
it displays unusually high symmetry, with only one type of Ru-Ru links, and uniform Ru-Cl-Ru bond
angles of $\approx$93$^{\circ}$.
By quantum chemical calculations in this particular honeycomb-lattice setting we find a very small
$J$, which yields a $K/J$ ratio as large as $\sim$100.
Interestingly, we also find that this is associated with vanishingly small $d$-shell trigonal
splittings, i.\,e., minimal departure from ideal $j_{\mathrm{eff}}\!=\!1/2$ moments.
This reconfirms RuCl$_3$ as a most promising platform for materializing the much sought-after Kitaev
spin-liquid phase and stimulates further experiments under strain and pressure.

\end{abstract}

\date\today
\maketitle

{\it Introduction.\,}
Sizable, bond-dependent Kitaev interactions \cite{Kitaev2006,Ir213_KH_jackeli_09} have been confirmed
by now in several honeycomb transition-metal (TM) oxides and halides.
The magnetic ground state however is in most of these systems ordered, due to residual isotropic
Heisenberg couplings, both nearest- \cite{Khaliullin_PRL_2010,Jiang_et_al} and farther-neighbor.
A main question is therefore which is the most suited chemical platform and set of structural
parameters (e.\,g., bond lengths and bond angles) maximizing the ratio between Kitaev and
Heisenberg exchange.

A distinct system in this context is RuCl$_3$ --- although it is ordered antiferromagnetically under
normal conditions, this antiferromagnetic phase lies in close proximity to the quantum spin liquid
\cite{Banerjee2016}.
In particular, the latter can be reached by applying a modest in-plane magnetic field \cite{Baek_2017,
Wang_2017,Zheng_2017}, which raises the question if strain or pressure could be used as well for
tuning the magnetic ground state.
Interestingly, a new crystalline phase has been recently identified under pressure of 1.26 GPa \cite{
rucl3_hp}.
Here we report {\it ab initio} quantum chemical results for the Ru-site multiplet structure and
effective intersite couplings in this recently discovered crystalline arrangement.
The computations reveal an unusually large $K/J$ ratio of $\sim$100, reconfirming RuCl$_3$ as one of
the most promising chemical settings for materializing the Kitaev spin-liquid ground state.
Additionally, we find a very peculiar inner structure of the effective moments, with $d$-shell trigonal
crystal-field splittings as low as 9 meV, 4--5 times less than in RuCl$_3$ under ambient pressure
\cite{Yadav2016,J3_winter_16,Warzanowski_et_al,Lebert_2020}.
This points to minimal departure from ideal $j_{\mathrm{eff}}\!=\!1/2$ moments \cite{Abragam1970,
Ir213_KH_jackeli_09}, through near cancellation of two different effects --- trigonal compression
of the ligand cage and anisotropic fields related to farther ions.
The apparent correlation between minimal departure from ideal, cubic-symmetry $j_{\mathrm{eff}}\!=\!1/2$
moments and maximized $K/J$ ratio is quite remarkable.
It seems to indicate that the nearest-neighbor $J$ is minimized in the case of degenerate $t_{2g}$
orbitals.

{\it Ru-site multiplet structure.\,}
The octahedral-cage ligand field splits the Ru 4$d$ levels into $e_g$ and $t_{2g}$ components, with
the latter lying at significantly lower energy; the large $t_{2g}$--$e_g$ splitting yields then a
$t^5_{2g}$ leading ground-state configuration.
With one hole ($s\!=\!1/2$) in the $t_{2g}$ sector ($l_{\mathrm{eff}}\!=\!1$), sizable spin-orbit 
coupling (SOC) provides a set of fully occupied $j_{\mathrm{eff}}\!=\!3/2$ and magnetically active
$j_{\mathrm{eff}}\!=\!1/2$ states.
For three-fold \cite{rucl3_hp} (or lower) site symmetry, the degeneracy of the $t_{2g}$ sublevels
is typically lifted and the $j_{\mathrm{eff}}\!=\!1/2$ and $j_{\mathrm{eff}}\!=\!3/2$ spin-orbit 
states may feature some degree of admixture.
To determine the Ru$^{3+}$ 4$d^5$ multiplet structure in $\alpha$-RuCl$_3$ at $p\!=\!1.26$ GPa, we
carried out quantum chemical computations using the {\sc molpro} suite of programs \cite{Molpro}
and crystallographic data as reported by Stahl {\it et al.}~\cite{rucl3_hp}.
A cluster consisting of one `central' RuCl$_6$ octahedron and the three in-plane adjacent octahedra
was designed for this purpose.
The crystalline environment was modeled as a large array of point charges which reproduces the Madelung
field within the cluster volume; to generate this point-charge embedding we employed the {\sc ewald}
program \cite{Klintenberg_et_al,Derenzo_et_al}.
The numerical investigation was initiated as a complete active space self-consistent field (CASSCF)
calculation \cite{olsen_bible,MCSCF_Molpro} with all five 4$d$ orbitals of the central Ru ion considered 
in the active orbital space.
Post-CASSCF correlation computations were carried out at the level of multireference configuration-interaction
(MRCI) with single and double excitations \cite{olsen_bible,MRCI_Molpro} out of the Ru 4$d$ and Cl 3$p$
orbitals of the central RuCl$_6$ octahedron.
SOCs were accounted for following the procedure described in Ref.~\cite{SOC_Molpro}
\footnote{
For the central Ru ion energy-consistent relativistic pseudopotentials (ECP28MDF) and Gaussian-type
valence basis sets (BSs) of effective quadruple-$\zeta$ quality (referred to as ECP28MDF-VTZ in the {\sc
molpro} library) \cite{4d_elements} were employed, whereas we used all-electron triple-$\zeta$ BSs
for the six Cl ligands of the central RuCl$_6$ octahedron \cite{Dunning_Cl}. 
The three adjacent TMs were represented as closed-shell Rh$^{3+}$ $t_{2g}^6$ species, using
relativistic pseudopotentials (Ru ECP28MDF) and (Ru ECP28MDF-VDZ) (4s4p3d)/[3s3p3d]
BSs for electrons in the 4th shell \cite{4d_elements}; the other 12 Cl ligands associated with 
the three adjacent TMs were described through minimal all-electron atomic-natural-orbital (ANO) 
BSs \cite{Pierloot1995}.}.

The Ru$^{3+}$ 4$d^5$ multiplet structure in the newly discovered crystalline phase of $\alpha$-RuCl$_3$
is depicted in Table~\ref{d5}, at three different levels of approximation: CASSCF, MRCI, and MRCI+SOC.
This allows to easily disentangle three different effects: crystal-field splittings, post-CASSCF
correlation-induced corrections, and of spin-orbit interactions.
As concerns the former, we find that the trigonal splitting within the $t_{2g}$ levels is tiny, 9 meV, 
4--5 times smaller than in $\alpha$-RuCl$_3$ at ambient pressure \cite{Yadav2016,J3_winter_16,
Warzanowski_et_al,Lebert_2020}. 
Having nearly degenerate $t_{2g}$ levels for sizable amount of trigonal distortion of the ligand cage
sounds odd at first.
What makes it happen is the presence of a competing effect --- trigonal fields related to the
anisotropy of the extended solid-state environment.
Mutual cancellation of those two yields cubic-like Ru-site multiplet structure.
Quite evident in Table \ref{d5} is also the presence of minimal splittings within the
$j_{\mathrm{eff}}\!=\!3/2$-like manifold once SOC is accounted for (last column), i.\,e., minimal
admixture of $j_{\mathrm{eff}}\!=\!1/2$ and $j_{\mathrm{eff}}\!=\!3/2$ spin-orbit states.
 
It is additionally seen that the post-CASSCF MRCI treatment yields sizable corrections to some of
the relative energies, the most substantial arising for the $^6\!A_{1g}$ crystal-field term.

\begin{figure}[t]
\includegraphics[width=0.95\columnwidth]{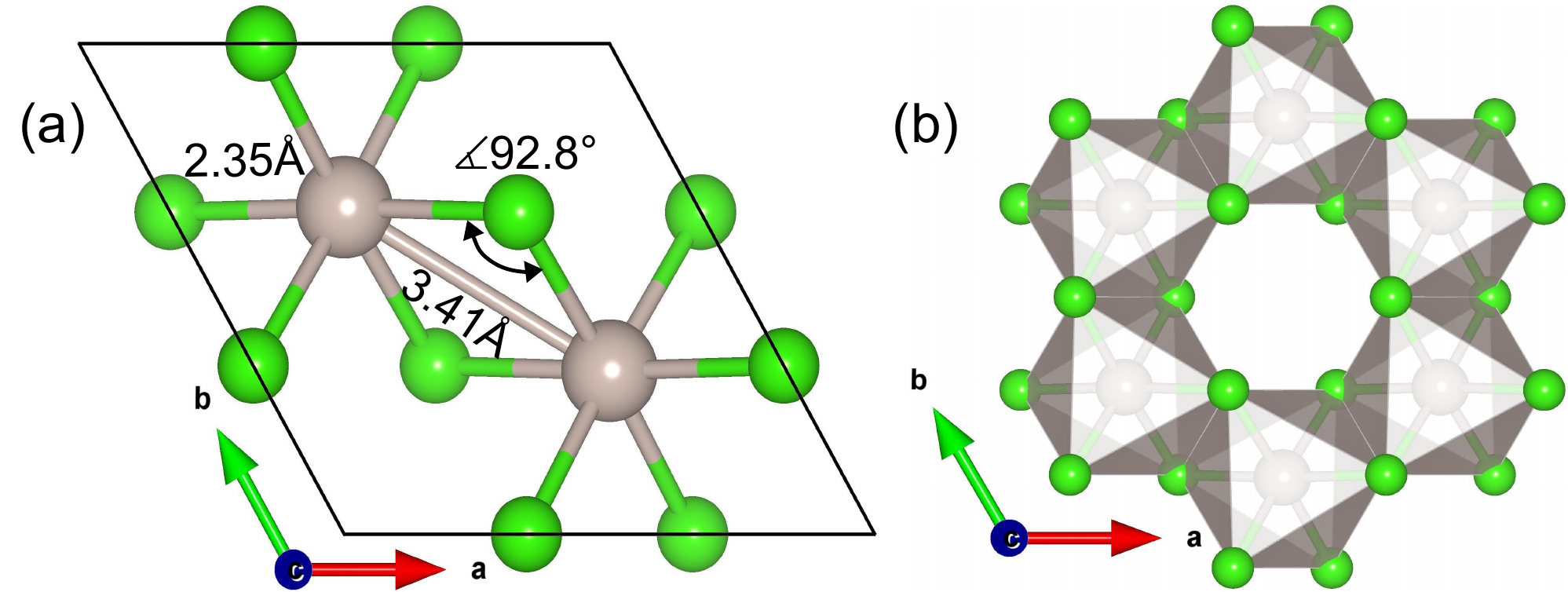}
\caption{
(a) Unit cell of $\alpha$-RuCl$_3$ for $p\!=\!1.26$ GPa \cite{rucl3_hp}.
(b) Hexagonal ring of edge-sharing RuCl$_6$ octahedra.
Grey and green spheres represent Ru and Cl ions, respectively.
}
\label{fig_1}
\end{figure}

\begin{table}[b]
\caption{
Ru$^{3+}$ $4d^5$ multiplet structure; all five $4d$ orbitals were considered in the active space.
Each value in the MRCI+SOC column indicates a Kramers doublet (KD);
for each of the $t_{2g}^4e_g^1$ crystal-field terms, only the lowest and highest KDs are shown.
Only the crystal-field terms enlisted in the table were included in the spin-orbit computation.
Notations corresponding to $O_h$ symmetry are used.
}
\begin{tabular}{llll}
\hline
\hline\\[-0.20cm]
Ru$^{3+}$ $4d^5$               &CASSCF     &MRCI   &MRCI\\
splittings (eV)                &           &       &+SOC\\
\hline
\\[-0.30cm]
$^2T_{2g}$ ($t_{2g}^5$)        &0          &0      &0   \\
                               &0.01       &0.01   &0.19\\
                               &0.01       &0.01   &0.20\\[0.10cm]
$^4T_{1g}$ ($t_{2g}^4e_g^1$)   &1.16       &1.32   &1.38\\
                               &1.17       &1.32   &$|$ \\
                               &1.17       &1.32   &1.49\\[0.10cm]
$^6\!A_{1g}$ ($t_{2g}^3e_g^2$) &1.15       &1.59   &1,78\,($\times 3$)\\[0.10cm]
$^4T_{2g}$ ($t_{2g}^4e_g^1$)   &1.84       &1.93   &2.06\\
                               &1.85       &1.94   &$|$ \\
                               &1.85       &1.94   &2.11\\
\hline
\hline
\end{tabular}
\label{d5}
\end{table}

{\it Intersite magnetic couplings for proximate $j_{\mathrm{eff}}\!=\!1/2$ moments.\,}
To obtain the intersite effective magnetic couplings, a cluster with two edge-sharing RuCl$_6$ octahedra 
in the central region was considered.
The four in-plane RuCl$_6$ octahedra coordinating this two-octahedra central unit were also explicitly
included in the quantum chemical computations but using more compact basis sets
\footnote{
We employed relativistic pseudopotentials (ECP28MDF) and BSs (ECP28MDF-VTZ) as also used in the single-octahedron 
computations \cite{4d_elements} for the central Ru species.
All-electron BSs of quintuple-$\zeta$ quality were utilized for the two bridging ligands \cite{Dunning_Cl}
and of triple-$\zeta$ quality for the remaining eight Cl anions \cite{Dunning_Cl} linked to the two octahedra
of the reference unit.
The four adjacent cations were represented as closed-shell Rh$^{3+}$ $t_{2g}^6$ species, using the same
pseudopotentials (Ru ECP28MDF) and BSs (Ru ECP28MDF-VDZ [3s3p3d]) \cite{4d_elements} considered for the
single-octahedron computations;
the outer 16 Cl ligands associated with the four adjacent octahedra were described through minimal ANO
BSs \cite{Pierloot1995}.}.
CASSCF computations were carried out with six (Ru $t_{2g}$) valence orbitals and ten electrons as
active (abbreviated as (10e,6o) active space)
\footnote{
The $t_{2g}$ orbitals of adjacent cations were part of the inactive orbital space.}.
Subsequently, three other types of wave-functions were generated, using in each case the orbitals 
obtained from the (10e,6o) CASSCF calculation\,: 
(i) single-configuration (SC) $t_{2g}^5$--$t_{2g}^5$ (i.\,e., the $t_{2g}^4$--$t_{2g}^6$ and
$t_{2g}^6$--$t_{2g}^4$ configurations which were accounted for in the initial (10e,6o) CASSCF 
were excluded in this case by imposing appropriate orbital-occupation restrictions),
(ii) (22e,12o) complete active space configuration-interaction (CASCI) wave-functions, as full
configuration-interaction expansions within the space defined by the Ru $t_{2g}$ and bridging-Cl
3$p$ orbitals,
and
(iii) MRCI wave-functions having the (10e,6o) CASSCF as kernel and accounting for single and double
excitations out of the central-unit Ru $t_{2g}$ and bridging-Cl 3$p$ orbitals.
By comparing data at these different levels of approximation, it is possible to draw conclusions on
the role of various (super)exchange mechanisms.

The CASSCF optimization was performed for the lowest nine singlet and lowest nine triplet states
associated with the (10e,6o) setting.
Those were the states for which SOCs were further accounted for \cite{SOC_Molpro}, either at SC,
CASSCF, CASCI, or MRCI level, which yields in each case a number of 36 spin-orbit states.

Only one type of Ru-Ru link is present in $\alpha$-RuCl$_3$ at $p\!=\!1.26$ GPa.
A unit of two nearest-neighbor octahedra exhibits $C_{2h}$ point-group symmetry, implying a 
generalized bilinear effective spin Hamiltonian of the following form for a pair of adjacent
1/2-pseudospins ${\bf{\tilde{S}}}_i$ and ${\bf{\tilde{S}}}_j$\,:
\begin{equation}
\label{eqn:Hamil}
\mathcal{H}_{ij}^{(\gamma)} = J{\bf{\tilde{S}}}_i\cdot {\bf{\tilde{S}}}_j + \\
                              K\tilde{S}_i^{\gamma}\tilde{S}_j^{\gamma} + \\
                              \sum_{\alpha\neq\beta} \Gamma_{\alpha\beta} \\
                              (\tilde{S}_i^{\alpha}\tilde{S}_j^{\beta} +  \\
                              \tilde{S}_i^{\beta}\tilde{S}_j^{\alpha}).
\end{equation}
The $\Gamma_{\alpha\beta}$ coefficients denote the off-diagonal components of the 3$\times$3
symmetric-anisotropy exchange matrix;
$\alpha,\beta,\gamma\!\in\!\{x,y,z\}$.
An antisymmetric Dzyaloshinskii-Moriya coupling is not allowed, given the inversion center.

The lowest four spin-orbit eigenstates from the {\sc molpro} output (eigenvalues lower by $\sim$0.2
eV with respect to the eigenvalues of higher-lying excited states, as illustrated for example in
Table\;I) were mapped onto the eigenvectors of the effective spin Hamiltonian (\ref{eqn:Hamil}),
following the procedure described in Refs.~\cite{Bogdanov_et_al,Yadav2016}\,:
those four expectation values and the matrix elements of the Zeeman Hamiltonian in the basis of the
four lowest-energy spin-orbit eigenvectors are put in direct correspondence with the respective
eigenvalues and matrix elements of (\ref{eqn:Hamil}).
Having two of the states in the same irreducible representation of the $C_{2h}$ point group, such
one-to-one mapping translates into two possible sets of effective magnetic couplings.
The relevant array is chosen as the one whose $g$ factors fit the $g$ factors corresponding to
a single RuCl$_6$ $t_{2g}^5$ octahedron.
We used the standard coordinate frame usually employed in the literature, different from the rotated
frame employed in earlier quantum chemical studies \cite{Yadav2016,Nishimoto2016,Ravi_CS} that affects
the sign of $\Gamma$ (see also discussion in \cite{NaRuO2}).

Nearest-neighbor effective magnetic couplings as obtained at four different levels of theory (SOC
included) are depicted in Table \ref{tab:couplings}.
The most remarkable finding is the vanishingly small $J$ value in the spin-orbit MRCI computations,
which yields a $K$-$\Gamma$-$\Gamma'$ effective spin model for the nearest-neighbor magnetic 
interactions.
That this coincides with realizing nearly ideal $j_{\mathrm{eff}}\!=\!1/2$ moments at the TM sites
seems to be more than merely fortuitous, see also discussion in the next section.

The competition between ligand and `crystal' trigonal fields (i.\,e., between nearest-neighbor and
beyond-nearest-neighbor electrostatics) and possible important implications as concerns the overall
magnetic properties of a given system have been earlier discussed in relation to single-site effective
magnetic paramaters such as the single-ion anisotropy \cite{Os227} --- in particular, it is in
principle possible to revert the sign of the latter by modifying the amount of ligand-cage trigonal
distortion \cite{Os227}.
Finding that this applies as well to intersite effective interaction parameters (i.\,e., the Heisenberg
$J$) is novel.

For comparison, the MRCI nearest-neighbor couplings in RuCl$_3$ at ambient pressure are {\it{K}}=--5.6,
{\it{J}}=1.2, $\Gamma$=1.2, and $\Gamma'$=--0.7 (meV) \cite{Yadav2016}.
The Heisenberg $J$ being sizable at ambient pressure, the $K/J$ ratio is much smaller than for the
set of parameters provided in Table~\ref{tab:couplings}.
This is realized for somewhat stronger trigonal compression of the Cl$_6$ polyhedron and additional
small distortions that actually lower the Ru-site point-group symmetry to less than trigonal.

Also notable is the fact that $K$ is basically the same at three different levels of approximation
(first column in Table~\ref{tab:couplings}):
SC (only the $t_{2g}^5$--$t_{2g}^5$ electron configuration considered), CASSCF (10e,6o) ($t_{2g}^5
$--$t_{2g}^5$ and $t_{2g}^4$--$t_{2g}^6$ configurations treated on the same footing), and CASCI
(22e,12o) (also excitations from the bridging-Cl 3$p$ to TM $t_{2g}$ orbitals taken into account).
It indicates that intersite TM\,$t_{2g}$\,$\rightarrow$\,TM\,$t_{2g}$ and ligand\,2$p$\,$\rightarrow$\,TM\,$t_{2g}$
excitations do not affect $K$.
What matters as concerns the size of the Kitaev coupling $K$ are 
(i) direct exchange, with a contribution of --1.75 meV, and
(ii) excitations to higher-lying states and so called dynamical correlation effects accounted for in
MRCI, with a contribution of --2 meV.

Anisotropic direct exchange as found in the SC calculation represents very interesting new physics, not
addressed so far in the literature.
Finding that nearly 50\% of the Kitaev effective coupling constant $K$ has to do with direct exchange 
and that the off-diagonal anisotropic coupling $\Gamma'$, which may give rise to spin-liquid ground
states by itself \cite{Ioannis_PRX}, comes $\sim$100\% from direct exchange (last column in Table\;\ref{tab:couplings})
obviously challenges present views and notions in Kitaev-Heisenberg quantum magnetism research and
superexchange theory. 
To provide additional reference points, we computed the isotropic direct exchange contribution \cite{
Anderson_1959} for holes in plaquette-plane TM orbitals having overlapping lobes along the Ru-Ru axis: 
this amounts to 25 meV, more than two times larger than direct exchange for the case of cuprate holes
in corner-sharing configuration of the ligand octahedra \cite{Martin_1993,VANOOSTEN1996}.
How exactly SOC and Coulomb interactions commix to yield large anisotropic direct exchange integrals
will be analyzed in detail elsewhere.
The important point however is that, at the $t_{2g}^5$--$t_{2g}^5$ SC level, there is a direct exchange
matrix element for each possible pair of holes --- $d_{xy}$-$d_{xy}$, $d_{xy}$-$d_{yz}$ etc.
SOC mixes up those different Slater determinants, and the resulting spin-orbit wave-functions are not
spin eigenstates.
The `spin-orbit' level structure can be reduced to an effective pseudospin model only by introducing
anisotropic direct exchange matrix elements (i.\,e., the SC values provided in Table\;\ref{tab:couplings}).

\begin{table}[t]
\caption{
Nearest-neighbor magnetic couplings (meV), 
results of spin-orbit calculations at
various 
levels of theory.
CASCI (22e,12o) stands for a full configuration-interaction within the space defined by the Ru 
$t_{2g}$ and bridging-Cl 3$p$ orbitals. 
The MRCI is performed having the (10e,6o) CASSCF wave-function as kernel.
}
\begin{tabular}{l c c c c}
\hline
\hline
\\[-0.25cm]
                 &{\it{K}}  &{\it{J}}  &$\Gamma_{xy}\!\equiv\!\Gamma$
                                                      &$\Gamma_{yz}\!=\!\Gamma_{zx}\!\equiv\!\Gamma'$\\

\hline
\\
[-0.22cm]
SC               &--1.75    &0.35      &--0.11        &0.42   \\[0.11cm]
CASSCF (10e,6o)  &--1.73    &--1.04    &0.89          &0.46   \\
CASCI (22e,12o)  &--1.72    &--1.04    &0.89          &0.47   \\[0.11cm]
MRCI             &--3.73    &--0.03    &1.62          &0.45   \\
\hline
\hline
\end{tabular}
\label{tab:couplings}
\end{table}

{\it Trigonal splittings in `213' iridate structures.\,}
Spotting this particular RuCl$_3$ crystalline arrangement, where
(i) the effects of ligand-cage trigonal compression and of farther-surrounding trigonal fields cancel
out each other,
(ii) the Ru-site $t_{2g}$ levels are consequently degenerate (or nearly degenerate), such that close to ideal
$j_{\mathrm{eff}}\!=\!1/2$ moments are realized, and
(iii) the intersite isotropic Heisenberg interaction approaches zero,
raises the question of whether an equivalent sweet spot can be identified in related Kitaev-Heisenberg
quantum magnets, e.\,g, in Ir-oxide honeycomb compounds.

Interestingly, that the Heisenberg $J$ changes its sign and therefore reaches a point where it simply
vanishes has been already pointed out, for both `213' hypothetical iridate structures \cite{Nishimoto2016}
and H$_3$LiIr$_2$O$_6$ \cite{Ravi_CS}.
Such a situation is achieved in iridates for ligand-cage trigonal squeezing providing Ir-O-Ir bond
angles of $\approx$98$^{\circ}$ \cite{Nishimoto2016,Ravi_CS} but an analysis of the on-site multiplet
spectra was not performed in those specific iridate crystalline settings.

To verify this important aspect, we carried out additional quantum chemical computations for `iridate'
clusters having only one octahedron as central region,
i.\,e., CASSCF calculations for a a cluster consisting of one central IrO$_6$ octahedron and three
adjacent octahedra in idealized Na$_2$IrO$_3$ setting with TM-ligand-TM bond angles of 98$^{\circ}$
\footnote{
For the central Ir ion relativistic pseudopotentials (ECP60MDF) and BSs of effective
quadruple-$\zeta$ quality (ECP60MDF-VTZ) \cite{Figgen_et_al} were used, 
whereas we applied all-electron triple-$\zeta$ BSs \cite{Dunning} for the six O 
ligands of the central IrO$_6$ octahedron. The three adjacent TMs were 
represented as closed-shell Pt$^{4+}$ $t^6_{2g}$ species, using relativistic 
pseudopotentials (Ir ECP60MDF) and (Ir ECP60MDF-VDZ) (4s4p3d)/[3s3p3d] BSs
\cite{Figgen_et_al}. The other 12 O ligands associated with the three
adjacent TM sites were described through minimal all-electron 
ANO BSs \cite{Pierloot1995}. Large-core 
pseudopotentials were employed for the 18 adjacent Na cations \cite{Fuentealba_Na}.}.
The outcome of this numerical test is rewarding:
also in the iridate system, a vanishing Heisenberg $J$ \cite{Nishimoto2016} is associated with
vanishing $d$-shell trigonal splittings, i.\,e., minor deviation from pristine $j_{\mathrm{eff}}\!=\!1/2$
states.
In particular, without accounting for SOC, we find a trigonal splitting of only 25 meV within the Ir
$t_{2g}$ levels, to be compared with a spin-orbit coupling constant of 400--500 meV for Ir ions.
That the near cancellation of ligand and `crystal' trigonal fields occurs for stronger trigonal
compression of the ligand cage ($\approx$98$^{\circ}$ vs $\approx$93$^{\circ}$ TM-ligand-TM bond
angles) has to do with the larger effective charges in iridium oxides (formally, Ir$^{4+}$ vs
Ru$^{3+}$ magnetic sites and O$^{2-}$ vs Cl$^{-}$ ligands).
Notably, various Kitaev-Heisenberg superexchange models do assume (for simplicity) degenerate $t_{2g}$
levels but not distorted TM-ligand-TM superexchange paths with bond angles away from 90$^{\circ}$.

{\it Conclusions.\,}
In spite of being central figure in nowadays research in quantum magnetism, textbook
$j_{\mathrm{eff}}\!=\!1/2$ spin-orbit ground states \cite{Abragam1970} are rarely found in solids
\cite{Ir2116_grueninger_2019,Ir2116_clancy_2020,Ir2116_tsirlin_2021}.
Here we show that nearly ideal $j_{\mathrm{eff}}\!=\!1/2$ moments are realized in a recently
reported crystalline phase of RuCl$_3$, identified under a pressure of 1.26 GPa \cite{rucl3_hp}.
In particular, we compute a vanishingly small trigonal splitting within the TM $t_{2g}$ valence 
subshell in this crystallographic setting.
Remarkably, this occurs in the presence of sizable trigonal squeezing of the ligand cages --- it
turns out that the effect of the latter is counterbalanced by trigonal fields having to do with 
the more distant crystalline surroundings.
Moreover, the nearly ideal $j_{\mathrm{eff}}\!=\!1/2$ character of the pseudospins is associated
with maximized $K/J$ ratio for the intersite magnetic interactions, through a vanishingly small 
value of the nearest-neighbor Heisenberg $J$.
The apparent correlation between these two features --- neat $j_{\mathrm{eff}}\!=\!1/2$ moments
and maximized $K/J$ ratio --- deserves careful further investigation, for instance, clarifying how
different (super)exchange mechanisms cancel out each other for degenerate on-site $t_{2g}$ orbital
energies but nevertheless distorted TM-ligand-TM paths when it comes to isotropic exchange.
Last but not least, we point out the important role of direct exchange \cite{Anderson_1959} in the
anisotropic spin interaction plot;
curiously, direct exchange contributions have been so far neglected in $K$-$J$-$\Gamma$-$\Gamma'$
exchange models.

 \ 

{\it Acknowledgments.\,}
P.\,B. and L.\,H. acknowledge financial support from the German Research Foundation (Deutsche
Forschungsgemeinschaft, DFG), Project No.~441216021, and technical assistance from U.~Nitzsche.
Q.\,S. and J.\,G. thank the DFG (SFB 1143, project-id 247310070) and the W\"urzburg-Dresden Cluster 
of Excellence on Complexity and Topology in Quantum Matter-ct.qmat (EXC 2147, project-id 390858490)
for financial support. 
We thank S.~Nishimoto and G.~Khaliullin for insightful discussions.

\bibliography{refs_feb01}

\end{document}